# Development of QoS methods in the information networks with fractal traffic

Yousef Daradkeh Eid MUBARAK[1], Lyudmyla KIRICHENKO[2,3], Tamara RADIVILOVA[2,3]
[1]*Stefan cel Mare University of Suceava, 720229, Romania*
[2,3]*Kharkiv National University of Radioelectronics, 61166, Ukraine*
*corresponding.author@mailserver.com*

*Abstract*— The paper discusses actual task of ensuring the quality of services in information networks with fractal traffic. The generalized approach to traffic management and quality of service based on the account of multifractal properties of the network traffic is proposed. To describe the multifractal traffic properties, it is proposed to use the Hurst exponent, the range of generalized Hurst exponent and coefficient of variation. Methods of preventing of network overload in communication node, routing cost calculation and load balancing, which based on fractal properties of traffic are presented. The results of simulation have shown that the joint use of the proposed methods can significantly improve the quality of service network.

## I. INTRODUCTION

At present along with the constant increase in data rates in telecommunications the proportion of interactive traffic which is extremely sensitive to the parameters of transportation is increasing. Task of ensuring QoS is becoming increasingly important, and its solution requires that the network tools were implemented, providing the acceptable quality of service [1].

QoS methods use different tools to reduce the negative effects of packets staying in queues whereas maintaining the positive role of queues. A set of tools is wide enough. [2]. Most of them take into account and use the existence of different types of traffic in the network in the sense that each traffic type has different performance requirements for the productivity and reliability of the network.

QoS methods use the queuing with different services algorithms on the network devices [1-3]. QoS methods are based on a gentle redistribution of the available bandwidth between the various types of traffic according to application requirements. It is obvious that these methods complicate the network devices, as it means the necessity to know the requirements of all traffic classes, to be able to classify them and distribute network bandwidth between them.

Traffic engineering methods are closely related to QoS techniques. Traffic engineering methods use change of traffic routes according to the actual channel load capacity [3]. According to the engineering methods transmission routes are managed so as to ensure balanced load of all network resources and avoid overloading communication devices and the formation of long queues. Therefore the effective modes of information exchange in the network are switching techniques, routing, and information flows control.

Multiple studies of processes in information networks have shown that the realizations of network traffic possess scale invariance property: self-similarity (fractality). For self-similar traffic the fast overload of nodes at low utilization rates is possible. It is especially evident if the network resources have been designed to load with the classical distribution flows. Investigation of self-similar traffic properties in telecommunication networks were carried out by such scientists as the W.E. Leland, M.S. Taqqu, W. Willinger, D.V. Wilson, O.I. Shelukhin. Self-similar properties were discovered in the local and global networks, particularly traffic Ethernet, ATM, applications TCP, IP, VoIP and video streams [4]. The reason for this effect consist in the features of the distribution of files on servers, their sizes, the typical behavior of users. It was found that data flows which don't have self-similarity initially after processing at the nodes and active network elements begin to show evident signs of self-similarity. The presence of self-similarity properties in information flows transmitted by customers have great influence on the performance of distributed systems. An especially important role it has for services, providing the transmission of multimedia and real-time traffic.

Thus, the task of development of fractal traffic management techniques is actual. The work purpose is to develop the generalized approach to management of self-similar traffic in a distributed system, which includes the management of channel capacity, buffering, routing and load balancing and are based on the analysis of fractal properties of the data flows and load of system resources.

## II. SELF-SIMILAR AND MULTIFRACTAL DATA FLOWS

Self-similarity of stochastic processes preserve the distribution laws when changing the time scale. Stochastic process X(t) is self-similar of index H, 0<H<1, if for any value a>0 processes and $a^{-H}X(at)$ have same finite-dimensional distributions:

$$\text{Law}\{a^{-H}X(at)\} = \text{Law}\{X(t)\} \quad (1)$$

Index H is called Hurst exponent. It is measure of self-similarity or measure of long-range dependence of process [5,6]. For values 0.5<H<1 time series demonstrates persistent behaviour. In other words, if the time series increases (decreases) in a prior period of time, then this trend will be continued for the same time in future. The value H=0.5 indicates the independence (the absence of any memory about the past) of values of time series. The interval 0<H<0.5 corresponds to antipersistent time series: if a system demonstrates growth in a prior period of time, then it is likely to fall in the next period.

The moments of the self-similar random process can be expressed as $E[|X(t)^q|]=C(q)t^{qH}$, where the quantity $C(q)$ is some deterministic function.

In contrast to the self-similar processes (1) multifractal random processes have more complex scaling behavior:

$$\text{Law}\{X(at)\} = \text{Law}\{\mathcal{M}(a) \cdot X(t)\}, \ a > 0, \quad (2)$$

where $\mathcal{M}(a)$ is random function that independent of X(t). In case of self-similar process $\mathcal{M}(a) = a^H$.

Multifractal traffic can be defined as extension of self-similar traffic by accounting for scalable properties of the statistical characteristics of second and higher orders. For multifractal processes the following relation holds:

$$E\left[|X(t)|^q\right] = c(q) \cdot t^{qh(q)}, \quad (3)$$

where $c(q)$ is some deterministic function, $h(q)$ is generalized Hurst exponent, which is generally non-linear function. Value at q=2 is the degree of self-similarity H. Generalized Hurst exponent of monofractal process does not depend on the parameter q: $h(q)=H$ [6].

Multifractal objects are statistically heterogeneous self-similar objects. For multifractal time series statistical heterogeneity of the object is reflected in the heterogeneous distribution of data, i.e. the heavy tails of the probability density function. Multifractal traffic has a special structure, preserving on many scales: some amount of very large bursts are always present in the realization with relatively small average level of traffic.

To characterize multifractal traffic it has been proposed to use the following parameters: intensity (average) of traffic $\lambda$, Hurst exponent H which defines the degree of long-range dependence (in particular, this means that for high traffic values most likely also high ones follow) and the range of the generalized Hurst exponent $\Delta h=h(q1)-h(q2)$, q1=-5, q2=5. Generalized Hurst exponent of monofractal process $h(q)=H$ and in this case $\Delta h=0$. The greater the heterogeneity of the process, i.e. more bursts are present in the traffic, the greater the range $\Delta h$.

The heaviness of the distribution tails corresponds to the degree of burstability. The coefficient of variation $\sigma_{var}(T)=\sigma(T)/M(T)$ can be considered as simple quantitative characteristic of tail of the traffic distribution T. In [7,8] it was shown that the increase (decrease) in the range of generalized Hurst exponent $\Delta h$ of multifractal time series corresponds to the increase (decrease) in value $\sigma_{var}$. Therefore, $\sigma_{var}$ can be used as the simplest characteristic of multifractal properties of traffic.

III. MODEL OF TRANSIT TRAFFIC IN COMMUNICATION NODE

Modern communication nodes are multitask, multiuser systems with multiple network interfaces, so they can transmit and receive multiple packets at the same time (depending on the amount of network interfaces on the switch) [9]. For this reason, the frequency of sending packets does not depend on jitter and packets size, and it can be represented with accuracy deterministic quantity sufficient for numerical calculations [1,5].

Model system of network traffic processing is presented as node LB with managed productivity or speed of data processing (number of packets per unit time), buffers specified size where the node place the traffic by one step (unit time), the pusher mechanism and the return [10,11].

Based on the above, the node can be represented as queuing system, which in each unit time receives the input data that come during this system work period, processes them and sends [12].

The node LB received traffic of several independent flows at each moment $t \epsilon T$. Traffic belong to the $qs \epsilon QS$ class of service which have to be delivered to i-th node $Serv_i$ for processing, without exceeding defined maximum permitted delay values $\tau_{qs}$ and maximum percentage of losses $l_{qs}$ depending on current nodes $Serv_i$ load and the actual channel capacity at moment of time.

Traffic has a lot of characteristics $V=\{\lambda,h,\mu\}$, where $\lambda=[\lambda_1,\lambda_2,...,\lambda_{Kn}]$ are intensity of multifractal applications flows (packages); $h=[H,h(q),\Delta h,\sigma_{var}]$, where $h(q)$ is selective value of the function of the generalized Hurst exponent, $h=h(2)$ is the value of Hurst exponent, $\Delta h=h(q_{min})-h(q_{max})$ is the range of values of the generalized Hurst exponent for the area of traffic, $\mu$ is the requirement of the request, $\sigma_{var}(T)$ is simple quantitative characteristic of tail of the traffic distribution. The requirement of the request is defined as a vector of required resources $\mu_{sr}=(CPU_{sr},Net_{sr},RAM_{sr})$ to complete the request. Each qs-th class of service corresponds to a set of vectors of requirement resources $\mu_{qs}=(CPU_r,Net_r,RAM_r)$.

The node LB and nodes $Serv_i$ are connected together by bilateral communication links with the maximum capacity $Link_{lk}=\{L_{lk}\}$, lk=1,2,..., that are divided into $k \epsilon K$ channels with bandwidth $Net_{lk}^k(t) = \{Net_{lk}^k\}$ at time t [12,13]. Each node $Serv_i$, i=1,2,... has the following characteristics: $CPU_i^{n_i}(T)$ is the average CPU utilization of each n processors of i-th server defined as the averaged CPU utilization during an observed period T; $RAM_i^{m_i}(T)$ is the average utilization of m memory of i-th server; $Net_i^{k_i}(T)$ is bandwidth of k channel of i-th server.

Each input flow is sent to a queue $Q_w$ of limited size. Queuing time is dependent on the class of service qs, i.e. the priority request is taken into account (the highest priority is the first). Whereas all priority service requests will not be processed, packages of other types stay in queue until the end of their lifetime. Newly received priority requests are dropping off the processing of non-priority ones and with a probability equal to one displace them in storage (if free waiting space have), or outside the system (if the storage is full). Packages displaced from service are join to queue of non-priority requirements and can be serviced after all the priority ones. Queues are separate, free space fully accessible for all newly received requests. Unlike typical priority queue system the considered system is equipped with probabilistic eject mechanism. The node LB, in accordance with a specified algorithm, extracts tasks from queues $Q_w$ and assign them to available resource of suitable nodes. [10,12]

To describe the mechanism of extrication of the network resources that occupied by the traffic at the end of transmission of traffic qs-th class of service (this occurs on the basis of data received from the routing protocol that supports communication on the available bandwidth and the available resources at the node (e.g., CSPF, SNMP)), let introduce the variable $\varepsilon_{Net_{lk}}^{qs,t_0}(t) = \{0,1\}$ that indicating that at the time t the traffic class qs-th stopped coming to node ($\varepsilon$=1), which was come to service at the time $t_0$ and had passed by the path $Net_i^{k_i}(t)$ to the node Serv$_i$. This variable contains all necessary data to determine the network resources that have to be extricated.

Node LB at moment t is characterized by coefficient of loss $X_{LB}^{qs}(t) \in X$, the average waiting time of package in the queue $T_{LB}^{qs}(t) \in T$. The variable $X_{LB}^{qs}(t) \in X$ is the percentage of loss traffic of qs-th class of service at node, that transmitted by the path $Net_{lk}^{k}(t)$ to the node Serv$_k$ at the moment t. It is assumed that the probability of package error in the path can be neglected and losses occur only in balancer because buffer overflows.

Following restrictions are applying to the coefficient of losses for all nodes in the network:

$$0 \leq X_{LB}^{qs}(t), \sum_{i=1}^{N} X_{LB}^{qs}(t) \leq l_{qs}, \qquad (4)$$

Thus, the restriction (4) show that the total loss for the traffic $\lambda_{Net_i}^{qs}(t)$ routed at time t, should not exceed the maximum permissible values for the class of service $l_{qs}$.

Loss is defined as the ratio of the discarded data to received data. Value $X_{LB}^{qs}(t) \to \min$ is subject to minimization. Restrictions imposed by the delay time are similar

$$0 \leq T_{LB}^{qs}(t), \sum_{i=1}^{N} T_{LB}^{qs}(t) \leq \tau_{qs}, \qquad (5)$$

where $T_{LB}^{qs}(t)$ is the average waiting time of package $qs$-th class of service in the queue on the i-th node. Performing this restriction helps to ensure that the delivery of packets does not exceed the maximum permissible values for a given class of service $\tau_{qs}$.

Thus, the operative dynamic management of the network nodes can be built based on estimation of current traffic parameters including the fractal properties and long-term dependence of information flows.

IV. METHOD OF BUFFERING AND CONTROL OF CHANNEL CAPACITY

The main tool for the study and forecasting networks with self-similar flows is simulation. In [11-13] simulation of channel load and study the formation of queues in the buffer for the realizations of fractal traffic were conducted. The results of simulation allowed to calculate the dependence of the buffer memory size $Q_w^{new} = f(Net_i^{k_i}, \lambda, H, \sigma_{var})$ on the values $Net_i^{k_i}(T)$ network bandwidth and parameters incoming data flow $\{\lambda,H,\sigma_{var}\}$. The calculated values of the buffer size $Q_w^{new}$ ensure the normal passing of traffic through the communication node (losses do not exceed the specified percentage). Similarly, the functional dependence of the capacity of the channel $Net_i^{newk_i} = \varphi(Q_w, \lambda, H, \sigma_{var})$ on the specified buffer size $Q_w$ and traffic parameters was obtained.

Functional dependencies $Q_w^{new} = f(Net_i^{k_i}, \lambda, H, \sigma_{var})$ and $Net_i^{newk_i} = \varphi(Q_w, \lambda, H, \sigma_{var})$ allow to determine maximum allowable load of the channel for the given size of the buffer memory and channel capacity. Calculating the value of the maximum allowable load in accordance with the obtained results, it is possible to predict it by traffic monitoring and avoid network overload by controlling the buffers and/or data flows. Forecasting the start of overload, can be allocated required size of the buffer memory $Q_w^{new}$ and/or channel capacity $Net_i^{newk_i}$. If the calculated buffer size $Q_w^{new}$ is larger than the existing $Q_w$, the system allocated the required size of buffer memory. We can also determine the required size of channel capacity from the received data. If the requested capacity size $Net_i^{newk_i}$ is larger than available $Net_i^{k_i}$, the system provides the requested resource, thus distributing the rest of the second channel capacity.

Method to control buffering and bandwidth switching node by monitoring of fractal traffic can be represented by the following steps:
– in the traffic that arrives at node, the fixed-length window is allocated;
– parameters $\{\lambda,H,\sigma_{var}\}$ for traffic in the selected window are estimated;
– according to $Q_w^{new} = f(Net_i^{k_i}, \lambda, H, \sigma_{var})$ and $Net_i^{newk_i} = \varphi(Q_w, \lambda, H, \sigma_{var})$ the size of the buffer memory $Q_w^{new}$ and/or the one of the channel capacity $Net_i^{newk_i}$ required to normal traffic passing through the communication node are predicted;
– if the estimated buffer memory $Q_w^{new}$ exceeds the size of the available buffer and/or the estimated size of the channel capacity $Net_i^{newk_i}$ is greater than available $Net_i^{k}$, it is necessary buffering and control of channel capacity;
– the next fixed-length window of traffic is analyzed etc.

This method can be used in network elements (switch, router and i.e.) for preventing of network overload. It allows to reduce a loss packets and increase channel utilization and network performance.

V. ROUTING METHOD IN THE NETWORK

Routes selection on the basis of separate flows is required to provide QoS, at that the different flows connecting one and the same pair of end points can be directed by various routes. In addition, in case of overload laid routes may be changed. The routing protocol based on link-state database calculates the shortest paths (Least Cost Routing) between the input edge router and all the rest ones [9,14,]. Consider the method of calculating cost of routing based on the fractal structure of traffic [7].

Assume that at each time $t$ traffic of intensity $\lambda_{Net_i}^{qs}(t)$ relating to one of the classes of service qs-th with requirements QoS, which correspond to the maximum delay value $\tau_q$ and the maximum percentage of loss $l_q$ is supplied to one of the routers [15,16]. All input traffic is divided at the flows of service classes so that to ensure the transmission requirements of all classes QS(t) in full [17]. Then, the channels set of QoS of traffic is described by $K = K(Net_{lk}^k, P_{lk}, L_k)$, where $Net_{lk}^k$ is channel bandwidth, $P_{lk} = \{p_{lk}^1, ..., p_{L_k}^k\}$ is allowable set of ways to the path $L_k$, that is defined for each traffic channel.

The value of routing cost $c_{lk}$ is assigned to the communication link lk and may depend on several parameters, particularly the speed, reliability and length. The cost of the path $p_{L_k}^k$ is equal to the sum of cost of communication lines: $C_{lk}^k = \sum_{lk \in p_{L_k}^k} c_{lk}$. If $Netx_{lk}^k(t)$ is bandwidth that is forwarded to the allowable path $p_{lk}^k$ of transmission channel of traffic $\lambda_{Net_i}^{qs}(t)$, then following relation holds $\sum_{t \in T; lk=1}^{L_k} Netx_{lk}^k(t) = Net_{lk}^k$, $\forall k \in K$, $\forall lk \in \{1, ..., L_k\}$.

The objective function that minimizes the cost of the routing cost to set of paths $P_{lk} = \{p_{lk}^1, ..., p_{L_k}^k\}$

$$\sum_{k \in K} \sum_{lk=1}^{L_k} C_{lk}^k Netx_{lk}^k(t) \to \min. \quad (6)$$

In [8] it was shown that at values H≥1 or at H>0.5 and simultaneously values $\sigma_{var} \geq 3$ (which roughly corresponds to the values $\Delta h > 1$) the amount of loss is greater than 5-10%. When passing traffic with strong fractal properties it needs to timely increase bandwidth of communication lines. To reflect the changes in the multifractal properties of flows, cost of paths $C_{lk}^k$ are updated in regular intervals and recalculated by the formula

$$Cnew_{lk}^k = \begin{cases} C_{lk}^k, & H \leq 0,5; \\ C_{lk}^k + (H - 0.5)C_0, & \\ 0.5 < H < 0.9, \sigma_{var} \leq 1; \\ C_{lk}^k + (H - 0.5)(\sigma_{var} - 1)C_0, & , \quad (7) \\ 0.5 < H < 0.9, 1 < \sigma_{var} < 3; \\ C_{lk}^k + C_0, & \\ H \geq 0.9 \text{ or } H > 0.5, \sigma_{var} \geq 3. \end{cases}$$

where $C_{lk}^k = \sum_{k \in p_{L_k}^k} c_{lk}$ is determined in accordance with the objective function (6), value $C_0$ is selected by the network administrator considering network topology. The routing algorithm is not changed (path cost $Cnew_{lk}^k = C_{lk}^k$) if the traffic has independent values (H=0.5) or has antipersistent properties (H<0.5). If 0.5<H<1 the value and the dispersion of data is small ($\sigma_{var} \leq 1$) the value $C_{lk}^k$ increases in proportion to the value of the Hurst exponent. If the value of the Hurst exponent 0.5<H<1 and dispersion is large (1<$\sigma_{var}$<3) the value $C_{lk}^k$ increases in proportion to both the characteristics. The cost with a maximum value $C_{lk}^k + C_0$ is obtained at H≥0.9 or persistent traffic (H>0.5) with a coefficient of variation $\sigma_{var} \geq 3$. After recalculating the value of all paths the announcement of the state of paths is sent between routers.

This method can be used to increase utilization channel by rerouting most important data flows to alternative low load channels.

VI. TRAFFIC BALANCING METHOD

Dynamic algorithm of load balancing based on monitoring of incoming traffic [18,19] and load of nodes system [20,21] is described in this paper. Step by step description of the dynamic load balancing algorithm with a modified feedback is present:

1) in the traffic that input to the node, the window of the fixed length T is selected;

2) value of the function of the generalized the Hurst exponent h(q), the value of Hurst exponent H=h(2) and the range of values of the generalized Hurst exponent $\Delta h = h(q_{min}) - h(q_{max})$ for the area of traffic are estimated in the dedicated window;

3) statistical data are collected and analyzed: the intensity of the incoming flows $\lambda_1, \lambda_2, ..., \lambda_{Kn}$ the load and imbalance of i-th nodes [22, 23] and system in general at a time t;

4) necessary amount of resources $\mu_{qs} = (CPU_i, Net_i, RAM_i)$ for each qs traffic classes are calculated on the basis of multifractal properties and intensity traffic;

5) calculations of flow distribution of network nodes are performed based on traffic classification and load and imbalance nodes and system; workload of nodes is forecasted using the data obtained in the next step;

6) traffic is balanced across the nodes, according to the load balancing algorithm considering each class of flows;

7) amount of underestimating resource $CPU_i^{n_i}$, $RAM_i^{m_i}$, $Net_i^{k_i}$ are balanced to nodes;

8) data about nodes and system utilization $CPU_i^{n_i}$, $RAM_i^{m_i}$, $Net_i^{k_i}$ are collected and are sent to the system load balancing to calculate the new flow distribution;

9) the next fixed-length window of traffic is analyzed.

The proposed load balancing method turned to providing a statistically uniform distribution of the load on nodes, high performance, capacity, fault tolerance (automatically detecting failures of nodes and redirecting the flow of data among the remaining) and low response time, the quantity of service information and losses.

## VII. SIMULATION OF THE DEVELOPED METHODS

To carry out simulation work of the proposed methods program modules were developed for the network simulator ns2, which allows simulating the operation of the information network by means of the proposed methods of Buffering and control of channel capacity, Routing method in the network and Traffic balancing method.

The input system receives the generated information flows having predetermined fractal properties with parameters similar to the real traffic. These flows form an additive multifractal traffic and sent to network, router control nodes, load balancer, which regulates the flow of tasks with the selected balancing policy.

To increase the fault tolerance in system, there is a Secondary Load Balancer, which performs the functions of a balancer, if the primary fails the load. Using this structure allows to manage and route traffic, load balance between the servers by the program components interact with each other.

In experiments by using model realizations traffic with predetermined properties a various network parameters can be determined in a variety of operation modes and loading.

In the constructed model network the following experiments were carried out:

- determination of the limiting capacities of different network fragments and dependence of packet loss on loading of individual gateways and external channels;
- determination of the response time of main nodes and servers in various modes of operation, in particular, when the network is loaded by 86-90%, which in real network is highly undesirable;
- determination of the necessary channels capacity to ensure a high level of quality of service;
- redistribution of traffic routes and redirection of the most critical information flows to less loaded alternative nodes;
- redistribution of tasks flows, that goes to servers, by using dynamic load balancing, taking into account the fractal properties of traffic and the imbalance of system resources for traffic of different classes.

During the simulation with method of buffering and control channel capacity, the dynamic control of channels capacity and sizes of buffer memory of nodes were performed (see Section IV).

During experiment, in the dynamic channel capacity control mode, based on the information obtained during the monitoring of self-similar traffic in the channels at the time t, a forecast of flows requirements in the necessary channel capacity $Net_i^{newk_i}$ were made at respective sections in the subsequent time interval t+Δ (ms)

Based on such prognostic evaluation of self-similar traffic critical to the time of transmission, the necessary resource of capacity $Net_i^{newk_i}$ was allocated, and the remaining traffic, respectively, the remaining resource for a time Δ.

Similarly, in the buffer size control mode of nodes, based on the information obtained in the process of monitoring self-similar traffic, buffer memory requirements of the node $Q_w^{new}$ were calculated for a time t+Δ. Based on this forecast, in the node required resources of buffer memory $Q_w^{new}$ (MB) were allocated for a while.

During the simulation with Routing method calculation of routing costs were carried out using the minimum cost criterion, with limiting the quality of service for the uniform use of traffic channels of different QoS classes in the case of multifractal flows (see Section V).

On the basis of information, obtained during the monitoring of fractal traffic in channels at the time t the route cost $Cnew_{lk}^{k}$ of channels from the sender's node to the recipient's node was calculated taking into account the required throughput $Net_{lk}^{k}(t)$. Based on this evaluation, the fractal traffic of high qsh class of service was routed along lowest cost paths taking into account traffic properties.

Most free communication line is allocated for traffic most critical to the delay. It was taken into account, that low-priority traffic in this communication line can wait a long time service. Accordingly, it is possible to multiplex low-priority traffic into separate flows so that its life time does not expire, and high-priority traffic is sent along other paths.

During the simulation with traffic balancing method, calculation of server load and system imbalance was performed (see Section VI). In network load balancing mode, based on the information obtained during the monitoring of self-similar traffic in channels at the time t, calculation of the system imbalance [19,22] and the necessary resources $CPU_i^{n_i}$, $RAM_i^{m_i}$, $Net_i^{k_i}$ for each qs traffic class was processed.

Subject to the limitations, least-loaded servers were selected. That is, for the most critical traffic by delay, the most suitable resources are allocated. At the same time taking into account that low-priority traffic can wait a long time for service.

In order to evaluate the effectiveness of proposed methods for managing the size of buffer memory, the dynamic distribution of the channel capacity and dynamic balancing in general, the results of simulation are analyzed.

During the experiment, the amount of data lost, channels load, value of jitter, imbalance of the system were measured. In table 1 shows parameters of network quality of service, that were obtained during the experiments.

With the same volume of transmitted information from sources to the receiver, the loss in the transmission of self-similar traffic is noticeably lower with simultaneous use of method of buffering and control of channel capacity, routing method in the network and traffic balancing method.

TABLE 1 – PARAMETERS OF QUALITY OF SERVICE

| Parameters | Method of buffering and control of channel capacity | Routing method in the network | Traffic balancing method | Using three methods |
|---|---|---|---|---|
| Average channel utilization | 0,7 | 0,6 | 0,73 | 0,5 |
| Average value of lost data, % | 1,86 | 1,92 | 1,8 | 1,7 |
| Average jitter, ms | 56 | 39 | 72 | 35 |
| System imbalance | 0,68 | 0,62 | 0,52 | 0,46 |

The results of the experiments realized and they brought the following effect: increasing the utilization of data transmission channels, by redirecting the most critical information flows to less loaded alternative channels, more efficient usage of system resources, improving the quality of service, and reduction of data loss.

VIII. CONCLUSION

The paper presents a generalized approach to the development of QoS methods in the information networks when fractal traffic transmitting. The most suitable tools have been selected for this, namely: management of channel capacity, buffering, routing and load balancing. The proposed methods are based on the multifractal properties of traffic passing through the network. To describe the multifractal traffic properties it is proposed to use the Hurst exponent which indicates degree of long-term dependence, the range of generalized Hurst exponent and coefficient of variation, characterizing the heterogeneity of the data flow.

Method of preventing of network overload, based on the calculation of the degree load of the communication channel by traffic monitoring is presented. This method can be used in the most congested parts of the network.

The developed method of routing cost calculation, which is based on the recalculated costs of ways depending on the fractal traffic properties is described. This method can be used in network routers.

Dynamic load balancing method, which takes into account the multifractal properties of traffic and value of the system imbalance, that provides a statistically uniform distribution of the load on the connection nodes is proposed. This method can be used for load balancing in distributed systems.

The proposed methods can be used in practice. Their application will optimize traffic management, allows for the dynamic distribute network resources, reduces costs by redirecting most critical information flows to less busy alternative channels and increases the degree of utilization of resources nodes.